\documentclass[aps,prb,reprint,twocolumn,showpacs,floatfix,superscriptaddress]{revtex4}
\usepackage{graphicx}
\usepackage{amssymb}
\usepackage{amsmath}
\usepackage{lmodern}
\usepackage{color}
\usepackage{hyperref}
\usepackage{caption}
\usepackage{epstopdf}
\begin{document}

\title{Electronic  structure of graphene: (nearly) free electrons bands  vs. tight-binding bands}

\author{E. Kogan}
\email{Eugene.Kogan@biu.ac.il}
\affiliation{Jack and Pearl Resnick Institute, Department of Physics, Bar-Ilan University, Ramat-Gan 52900, Israel}

\author{V. M. Silkin}
\email{vyacheslav.silkin@ehu.es}
\affiliation{Donostia International Physics Center (DIPC), Paseo de Manuel Lardizabal 4, E-20018 San Sebastian/Donostia, Spain,\\
Departamento de F\'{\i}sica de Materiales, Facultad de Ciencias Qu\'{\i}micas, UPV/EHU, Apartado 1072, E-20080 San Sebastian/Donostia, Spain \\
IKERBASQUE, Basque Foundation for Science, 48011 Bilbao, Spain }

\date{\today}

\begin{abstract}
In our previous paper (Phys. Rev. B {\bf 89}, 165430 (2014)) we have
found that in graphene, in distinction to the four occupied bands,
which can  be described by the simple tight-binding  model (TBM)
with four atomic orbitals per atom, the two lowest lying at the
$\Gamma$-point unoccupied bands  (one of them of a $\sigma$ type and
the other of a $\pi$ type) can not be described by such model. In
the present work we suggest a minimalistic model for these two
bands, based on (nearly) free electrons model (FEM), which correctly
describes the symmetry of these bands, their dispersion law and
their localization with respect to the graphene plane.
\end{abstract}


\maketitle

\section{Introduction}

In the course of the study of graphite and a graphite monolayer,
called graphene, understanding of the symmetries of the electron
dispersion law in graphene was of crucial importance. Actually, the
symmetry classification of the energy bands in graphene was
presented nearly 60 years ago by Lomer in his seminal paper
\cite{1}. Later the subject was analyzed by Slonczewski and Weiss
\cite{2}, Dresselhaus and Dresselhaus \cite{3},  Bassani and
Parravicini \cite{4}, Malard et al \cite{5}, Manes \cite{6} and
others. Note that the first attempt to  apply the TBM to the
calculation of the electrons dispersion law in graphene (graphite)
was done already in 1935 by
 F. Hund and B. Mrowka \cite{hund}.

 In our previous publication \cite{kogan}
we compared the classification of the electron bands in graphene,
obtained by group theory algebra in the framework of a tight-binding
model (TBM), with that calculated in a density-functional-theory
(DFT)  local-density approximation framework. Identification in the
DFT band structure of all eight energy bands (four valence and four
conduction bands) corresponding to the TBM-derived energy bands was
performed and the corresponding analysis was presented. The four
occupied (three $\sigma$ -like and one $\pi$ -like) and three
unoccupied (two $\sigma$ -like and one $\pi$ -like) bands given by
the DFT closely correspond to those predicted by the TBM, both by
their symmetry and their dispersion law. However, the two lowest
lying at the $\Gamma$-point unoccupied bands (one of them of a
$\sigma$ -like type and the other of a $\pi$-like one), were found
to be not of the TBM type in accord with the earlier publications
\cite{pobaprl83,pobaprl84,wegrprl08}.
Moreover, recently it was shown
that indeed these two states resemble the lowest-energy members of
the image-potential states series predicted to exist in vicinity of
the graphene monolayer when the potential shape on the vacuum side
is modified to a correct asymptotic $\sim -1/4z$ behavior
\cite{sizhprb09}. Indeed such unoccupied image-potential states at
the solid surfaces \cite{ecpejpc78} were a subject of extensive
theoretical and experimental investigation during several decades
\cite{dast92,farecp00,ecbessr04}. Recently image-potential states
were observed on graphene absorbed on different substrates
\cite{bobaprl10,zhhujpcm10,nifaprb12,arguprl12,nopoprb13,crruprl13,nifajpcm14},
whereas the two lowest-energy members of an image series of a pure
graphene were detected for graphene placed of a SiC surface
\cite{bosinjp10,taimprb14,shjoapl14}. Moreover a transformation of
such states in carbon nanotubes and fullerenes
\cite{fezhs08,zhfeacsn09} or an evolution of the lowest-energy
symmetric state into a widely discussed interlayer band
\cite{fahiprl83,stblprb00,cslinp05} in graphite upon increasing of
number of the graphene layers was studied
\cite{bosinjp10,taimprb14,codoss16}.

\begin{figure}[h]
\includegraphics[width= \columnwidth, trim=25 0 20 0, clip=true]{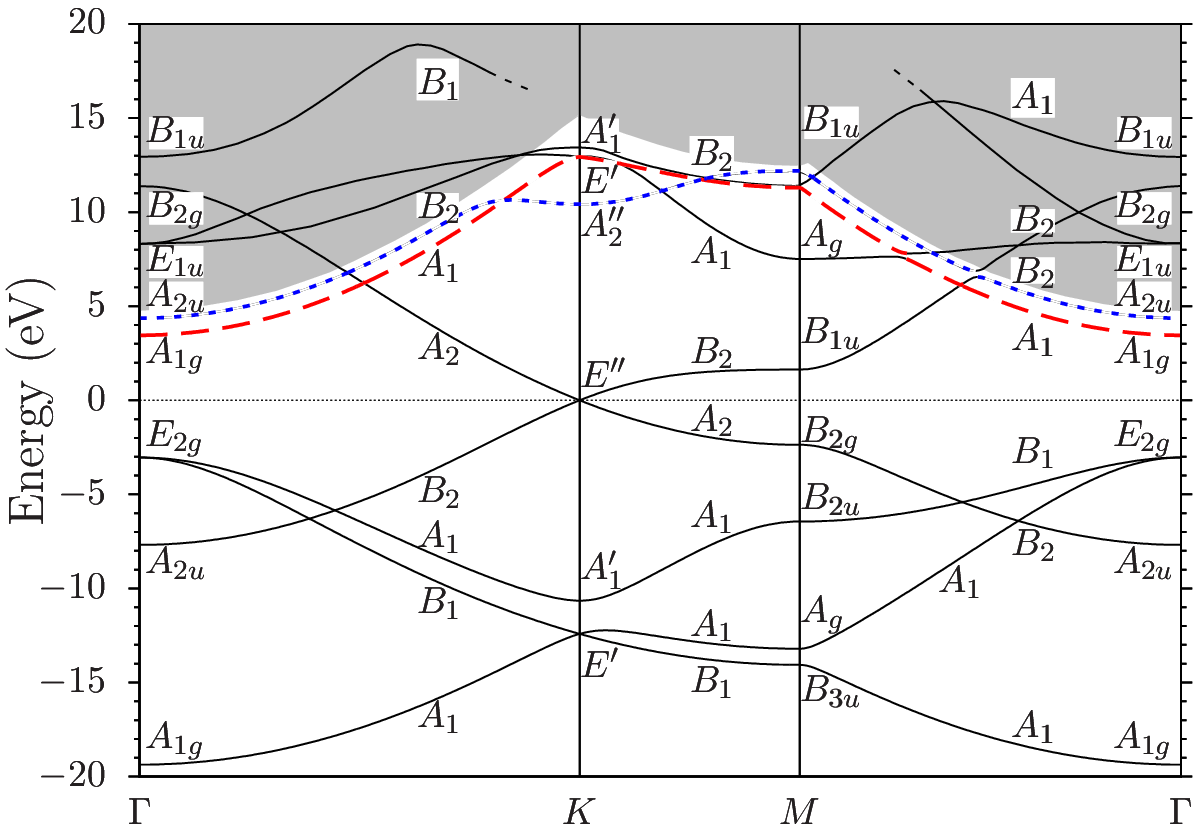}
\caption{\label{fig:bands}  (color online) Graphene band structure.
The dotted line shows the Fermi energy. The grey area corresponds to
the vacuum continuum states. The non-TBM bands of our interest are
plotted with long-dashed (red) and short-dashed (blue) lines. The
upper bands merge into scattering resonances of graphene
\cite{Nazarov-13,Wicki-16}, which is schematically shown with dashed
(black) endings of lines. }
\end{figure}

\begin{figure}[h]
\includegraphics[width= .8\columnwidth, trim=25 0 20 0, clip=true]{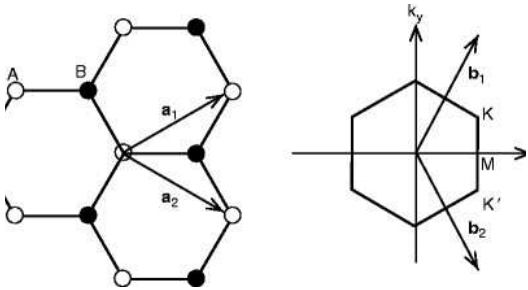}
\caption{\label{fig:bandsn}   Honeycomb lattice and its Brillouin
zone. }
\end{figure}

In Fig. \ref{fig:bands} we reproduce
 the results of the band structure calculations with symmetry labelling of the
occupied and the lowest lying unoccupied bands.  The TBM  bands are
plotted in solid lines, $\sigma$ non-TBM band - in long dashed,
$\pi$  non-TBM band - in dotted line.
To understand the symmetry
classification of the bands one should remember that the group of
wave vector ${\bf k}$  at the $\Gamma$ point is $D_{6h}$;
  at the $K$ point -- $D_{3h}$;  at the point $M$ -- $D_{2h}$.
The group of wave vector ${\bf k}$  at each of the lines
constituting triangle $\Gamma-K-M$  is $C_{2v}$
\cite{thomsen,dresselhaus}.
 Representations of the groups can be
found in the book by Landau and  Lifshitz \cite{landau}. Honeycomb
lattice and its Brillouine zone with the symmetry points are
presented on Fig. \ref{fig:bandsn}.
One of rotations $U_2$ for the $D_{6h}$ group is about the   direction $\Gamma - K$.  Rotation  $C_2^z$  for the $D_{2h}$ group is about the normal to graphene plane, rotation
 $C_2^x$  - about the $\Gamma - M$ line.
Reflection $\sigma_v$ for the $C_{2v}$ groups is relative to the plane of graphene.
The mathematical details of the symmetry analysis in TBM are presented in  Section \ref{TBM}.

\section{Free electrons model  bands}

Let us start this Section from explaining what do we mean by saying that a given electron band is not a tight-binding one.
 In our previous publication \cite{kogan} we have
pointed that the symmetry of the two  low lying unoccupied bands contradicts to the predictions of the tight-binding model (with the four atomic orbitals considered). We have also shown that their localization with respect to graphene plane is drastically different from that of occupied bands and also of $\pi^*$ band.   In the present Section we make another step in this direction.
We  point that both the symmetry and the dispersion law of these two  bands can be obtained in the framework of the model opposite
to the tight-binding model -- the (nearly) free electrons model.

The  difference between the TBM and non-TBM states
is evident from the Figures in our previous publication \cite{kogan} presenting electron density. The former
are localized in the vicinity of graphene plane within a distance of the order of graphene lattice constant $a$.
The latter are localized within a distance $\mu a$, where $\mu\gg 1$.
It means that
while calculating the dispersion law  with its typical energy scale $\hbar^2/ma^2\sim 1$ eV, for the  non-TBM bands in zero order approximation with respect to parameter $1/\mu$  we can ignore
the $z$ dependence of the wave functions.
In addition,
when looking on Fig. \ref{fig:bands}
one notices that the dispersion of the two non-TBM bands follow very closely the continuum bottom lines, which present the dispersion law of the free electrons, staying below them.

Hence we can formulate  a minimalistic model for the  non-TBM bands by considering them as (nearly) free electrons model (FEM) bands,
more specifically by presenting their wave functions  in the factorised form
\begin{eqnarray}
\psi_{\bf k}(x,y,z)=f_{\bf k}(z)\phi_{\bf k}(x,y),
\end{eqnarray}
where $\phi_{\bf k}(x,y)$ are (nearly) free electron wave functions corresponding to the continuum bottom, and the functions $f(z)$ are determined by the boundary conditions
\begin{eqnarray}
\lim_{z\to\pm\infty}f_{\bf k}(z))=0.
\end{eqnarray}
For the $\sigma$ band $f(z)$ is an even function, and for   the  $\pi$ band -- an odd one. The multiplier $f(z)$ would give a
first order with respect to parameter $1/\mu$ negative correction to the dispersion of the FEM bands, putting them slightly below the continuum bottom. However, for the symmetry analysis presented below, this multiplier is irrelevant, apart from the fact of its parity, the former distinguishing between $\sigma$ and $\pi$ band.

According to the nearly-free-electron model, the wave functions of the states inside the Brillouin zone are just plane waves. On the boundaries of the zone they are combinations
of small number of plane waves \cite{kittel}.
Thus, on the lines $\Gamma -K$ and $\Gamma -M$  $\phi_{\bf k}(x,y)=e^{i{\bf k}\cdot {\bf r}}$;
on the line $K -M$, $\phi_{\bf k}(x,y)$ is an arbitrary linear combinations of two plane waves: $e^{i{\bf k}\cdot {\bf r}}$ and the other one, corresponding to the equivalent point on the opposite
edge of the hexagon, forming the boundary of the Brillouin zone;
at the point $K$, $\phi_{\bf K}(x,y)$ is a linear combination of three plane waves
$e^{i{\bf K}_{1}\cdot {\bf r}},e^{i{\bf K}_{2}\cdot {\bf r}},e^{i{\bf K}_{3}\cdot {\bf r}}$
 corresponding to the three equivalent vertices of the hexagon ${\bf K}_{1,2,3}$.

Our assumption allows to   derive  the symmetry realized by each of the two bands from the symmetry of plane waves
with respect to rotations and reflections in the plane (and the symmetry of the function $f(z)$ with respect to reflection in the $z=0$ plane).

Thus we immediately obtain that  at the point $\Gamma$, the $\sigma$ band realizes representation $A_{1g}$, and  the $\pi$ -- representation $A_{2u}$.
At the lines $\Gamma -K$ and $\Gamma -M$, the $\sigma$ band realizes representation $A_{1}$, and  the $\pi$ -- representation $B_{2}$.

At the line $K-M$ the function $f(z)\times$ (sum of the exponents)
 realizes $A_1$ representation of $C_{2v}$ for even $f(z)$,
and $B_2$ representation for odd $f(z)$; the function
$f(z)\times$ (difference of the exponents) realizes $B_1$
representation for even $f(z)$,
and $A_2$ representation for odd $f(z)$.
One can expect that the two bands considered  correspond to symmetric combinations of the plane waves.

At the point $M$ the  function
$f(z)\times$ (sum of the exponents) realizes $A_g$
representation of $D_{2h}$ for even $f(z)$,
and $B_{1u}$ representation for odd $f(z)$;
the function $f(z)\times$ (difference of the exponents)
realizes $B_{3u}$ representation of $D_{2h}$ for even $f(z)$,
and $B_{2g}$ representation for odd $f(z)$. Again, one can expect that the two bands considered would correspond to symmetric combinations of the plane waves.

To expand the representation realized  by the plane waves at  the point $K$
we may use equation \cite{landau}
\begin{eqnarray}
\label{char}
a^{(\alpha)}=\frac{1}{g}\sum_{G}\chi(G)\chi^{\alpha}(G)^*,
\end{eqnarray}
where $a^{(\alpha)}$ shows how many times an irreducible representation $\alpha$ of the group (in our case $D_{3h}$) occurs in the expansion,
$g$ is the number of elements in the group,
$G$  is an arbitrary element of the group, $\chi(G)$ is the character of the element in the origional representation, and $\chi^{\alpha}(G)$ is the character of the element in the irreducible representation  $\alpha$.

Taking into account the table of the characters of the irreducible representations of the group $D_{3h}$ we get
that at the point $K$ the functions $f(z)\times\left\{e^{i{\bf K}_{1}\cdot {\bf r}},e^{i{\bf K}_{2}\cdot {\bf r}},e^{i{\bf K}_{3}\cdot {\bf r}}\right\}$
realize
$A_1'+E'$ representation of $D_{3h}$ for even $f(z)$,
and $A_2''+E''$ representation for odd $f(z)$ \cite{kogan1}. More specifically, according to the FEM \cite{kittel}, in the vicinity of the K point there are 3 $\sigma$ and 3 $\pi$ bands which are linear combinations  of the 3 plane waves, corresponding to the equivalent vertices of the Brillouine zone hexagon (multiplied by the appropriate $f(z)$ functions.
Judging by the band structure presented on Fig. \ref{fig:bands}, in each case two upper bands are high in the continuum, and the lower one is below the continuum bottom.

The  FEM $\sigma$ states in the vicinity of the K point are hybridized
with the TBM $\sigma$ states. Thus the $E'$ representation of
the unoccupied bands is   realized by  the states of both types \cite{kogan1,kogan} (see also the next Section).

On the other hand, the  FEM $\pi$ states in the vicinity of the K point have no TBM counterparts, to hybridize with.
However,
the lower FEM band in the vicinity of the K point  is repelled by the upper bands.
 (Formally, the upper bands exist all over the Brillouine zone, but they come close to the lower one only  in the vicinity of the $K$ point.)
This explains why the FEM $\pi$ band deviates from the bottom of the continuum there.

\section{TBM bands}
\label{TBM}

In our previous publications \cite{kogan1,kogan} we performed the
symmetry analysis of the electron bands in the framework of the TBM.
In this Section we want to expand and to substantiate this analysis,
which make it easier the comparison with the symmetry analysis for
the free electrons bands. In the beginning  we provide additional
(to what was presented in our previous publications
\cite{kogan1,kogan}) group theory arguments for the symmetry
assignment.

To explain the message of the rest of the Section, let us start from
asking a general question: How can we know what  the symmetry of a
given electron band, obtained in the process of DFT calculation, is?
One approach to answering this question is based on compare the way
the calculated bands (taking into account all of them
simultaneously) merge at the points $\Gamma$ and $K$ with the
predictions of the group theory, based on the symmetry of the
considered atomic orbitals and the graphene lattice. This is the way
we followed in our previous publications \cite{kogan1,kogan}. This
approach can give essential information but it has essential
limitations, apart from being heuristic. Even if we believe that it
describes correctly the symmetry at the merging point, it does not
always allow to tell which symmetry has the particular band at the
symmetry line. The second limitation is more severe. Because of the
absence of merging at the point $M$, we can make only a (un)educated
guess about the symmetry of a given band at that point.

To know for sure, what the symmetry of a given band is everywhere,
one has to know not only the dispersion law, but also the
wavefunction corresponding to the band (actually the density is
enough).  It turns out, that, whether at the point $M$, or at the
symmetry line, for a given band one has to choose between two
different representations. To make the choice, one has to find the
symmetry axis such, that one of the representation is even with
respect to reflection about the axis, and the other is odd, and
hence the wavefunction of the band, realizing the latter reflection,
is equal to zero at the axis. Such analysis proves  our previous
guesses \cite{kogan} about the symmetry of the bands.

Now let us recall the basics of TBM.
We look for the solution of the Schr\"{o}dinger equation as a linear combination of the functions
\begin{eqnarray}
\label{tb}
\psi_{\beta;{\bf k}}^j=\sum_{{\bf R}_j} e^{i{\bf k\cdot R}_j}\psi_{\beta}\left({\bf r}-{\bf R}_j\right),
\end{eqnarray}
where $\psi_\beta$ are atomic orbitals, $j=A,B$ labels the sub-lattices, and  ${\bf R}_j$ is the radius vector of an atom in the sublattice $j$.
Our TBM space includes four atomic orbitals: $|s,p>$. (Notice that we assume only symmetry of the basis functions with respect to rotations and reflections; the question how these functions  are
related to the atomic functions of the isolated carbon atom is irrelevant.)

The Hamiltonian of graphene being symmetric with respect to reflection in the graphene plane, the bands built from the $|p_z>$  orbitals decouple from those built from the $|s,p_x,p_y>$  orbitals. The former are odd with respect to reflection, the latter are even. In other words, the former form $\pi$  bands, and the latter form $\sigma$  bands.

A  symmetry transformation of the functions $ \psi_{\beta;{\bf k}}^j$ is a direct product of two transformations: the transformation of the sub-lattice functions $\phi^{A,B}_{{\bf k}}$, where
\begin{eqnarray}
\label{2}
\phi_{\bf k}^j=\sum_{{\bf R}_j} e^{i{\bf k\cdot R}_j},
\end{eqnarray}
and the transformation of the orbitals $\psi_{\beta}$. Thus the representations realized by the functions (\ref{tb}) will be the direct product of two representations.

\subsection{$\Gamma$ - point}

Let us start from the most symmetrical  point $\Gamma$.
The functions $\phi_{\bf 0}^{A,B}$ realize $A_{1g}+B_{1u}$ representation of the group $D_{6h}$.
The orbital $|s>$ realizes $A_{1g}$ representation of the group, $|p_z>$ realizes $A_{2u}$ representation,
the orbitals $|p_x,p_y>$ realize $E_{1u}$ representation.
The identitity
\begin{eqnarray}
\label{g1}
(A_{1g}+B_{1u})\times A_{1g}=A_{1g}+B_{1u}
\end{eqnarray}
specifies the  two $\sigma$ bands constructed from $|s>$ orbitals; the identity
\begin{eqnarray}
\label{g2}
(A_{1g}+B_{1u})\times A_{2u}=A_{2u}+B_{2g}
\end{eqnarray}
specifies the  two $\pi$ bands constructed from $|p_z>$ orbitals; the identity
\begin{eqnarray}
\label{g3}
(A_{1g}+B_{1u})\times E_{1u}=E_{1u}+E_{2g}
\end{eqnarray}
shows that  there are two merging points of the $\sigma$ bands constructed from $|p_{x,y}>$ orbitals.

Eqs. (\ref{g1}) - (\ref{g3}) explain why  the band  $A$ lies below the
corresponding   $B$ band, and   the degenerate  bands corresponding to  $E_1$ representation are above those corresponding to the   $E_2$ representation.  In fact, the symmetry of the band(s) with respect to exchange of the lattice sites (the higher symmetry corresponding to the lower band
 \cite{kittel}) can be deduced from the
symmetry of the representation realized by $\phi_{\bf k}^j$.

\subsection{$K$ - point}

Now consider the point $K$.
The functions $\phi_{\bf K}^{A,B}$ realize $E'$ representation of the group $D_{3h}$.
The orbitals $|s>$ and $|p_z>$ realize $A_1'$ and
$A_2''$
representations respectively,  the orbitals $|p_x,p_y>$ realize
$E'$ representation.
The product of the representations is expanded as
\begin{eqnarray}
\begin{array}{l}
E'\times A_1'=E'\\
E'\times A_2''=E''\\
E'\times E'=A_{1}'+E'+A_2'.\end{array}
\end{eqnarray}
Notice that $A_{1}'+E'$ is the symmetric product of the representation on itself (including, of course, the absolutely
symmetric representation  $A_{1}$ and $A_2'$ is the antisymmetric product \cite{landau}.)
Taking into account these identities,
we obtain
a band realizing $A_1'$ representation and that realizing $A_2'$
representation, constructed from $|p_x,p_y>$ orbitals,
two merging points realizing $E'$ representations,
each of them constructed from $|s,p_x,p_y>$ orbitals and  the merging point realizing $E''$ representation,
constructed from $|p_z>$ orbitals.

\begin{figure}[h]
%\centering
\hskip -.5cm
  \begin{minipage}[b]{0.25\textwidth}
  \includegraphics[width= 1\textwidth]{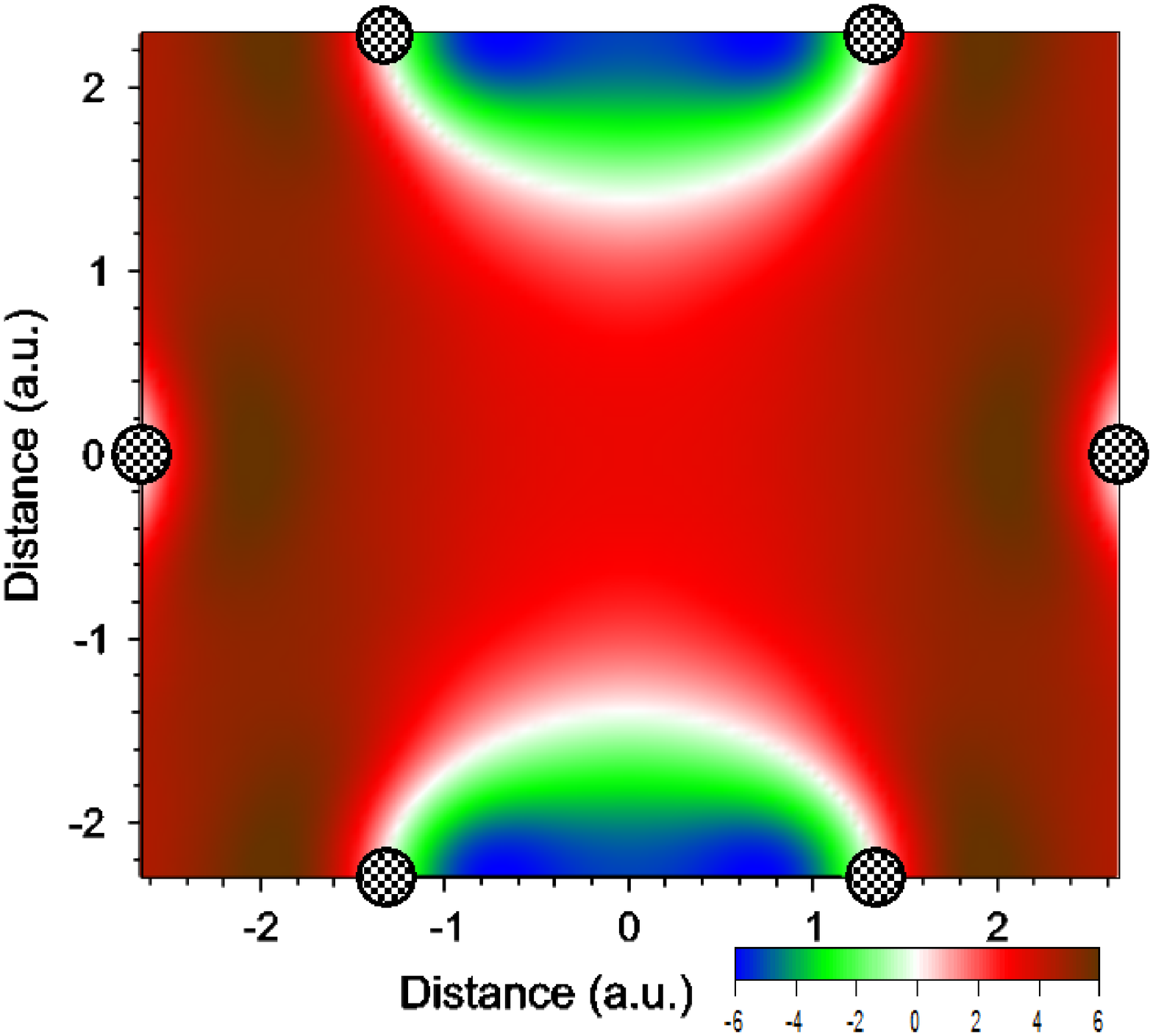}
\end{minipage}
  \hfill
  \begin{minipage}[b]{0.25\textwidth}
\includegraphics[width= 1\textwidth]{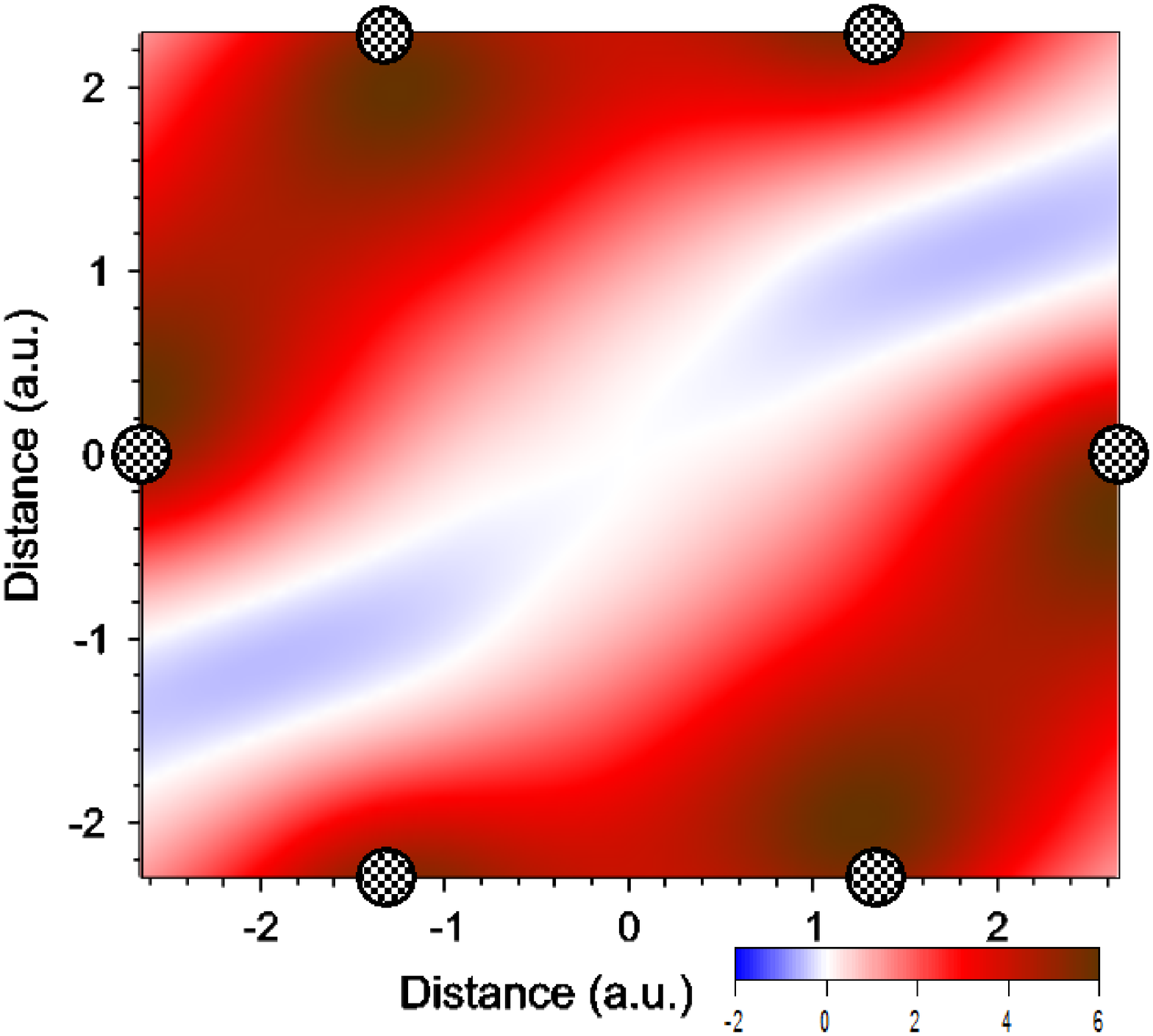}
\end{minipage}
\caption{\label{fig:3k}  (color online)  Real (left) and imaginary (right) parts of the wave function (in arbitrary
units) at the $z = 0$ plane for the third lowest band at $K$ point. }
\end{figure}

\begin{figure}[h]
%\centering
\hskip -.5cm
  \begin{minipage}[b]{0.25\textwidth}
  \includegraphics[width= 1\textwidth]{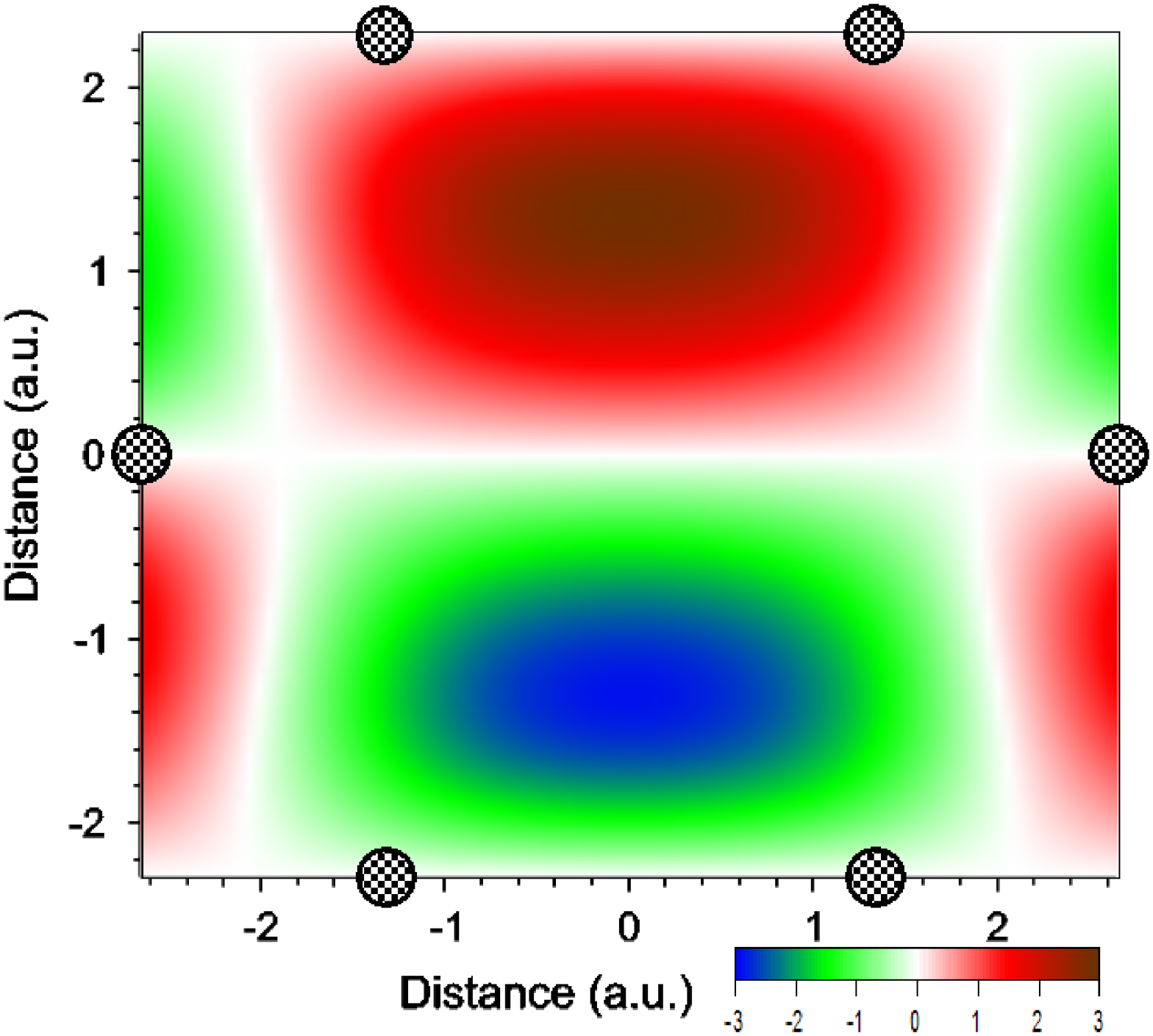}
\end{minipage}
  \hfill
  \begin{minipage}[b]{0.25\textwidth}
\includegraphics[width= 1\textwidth]{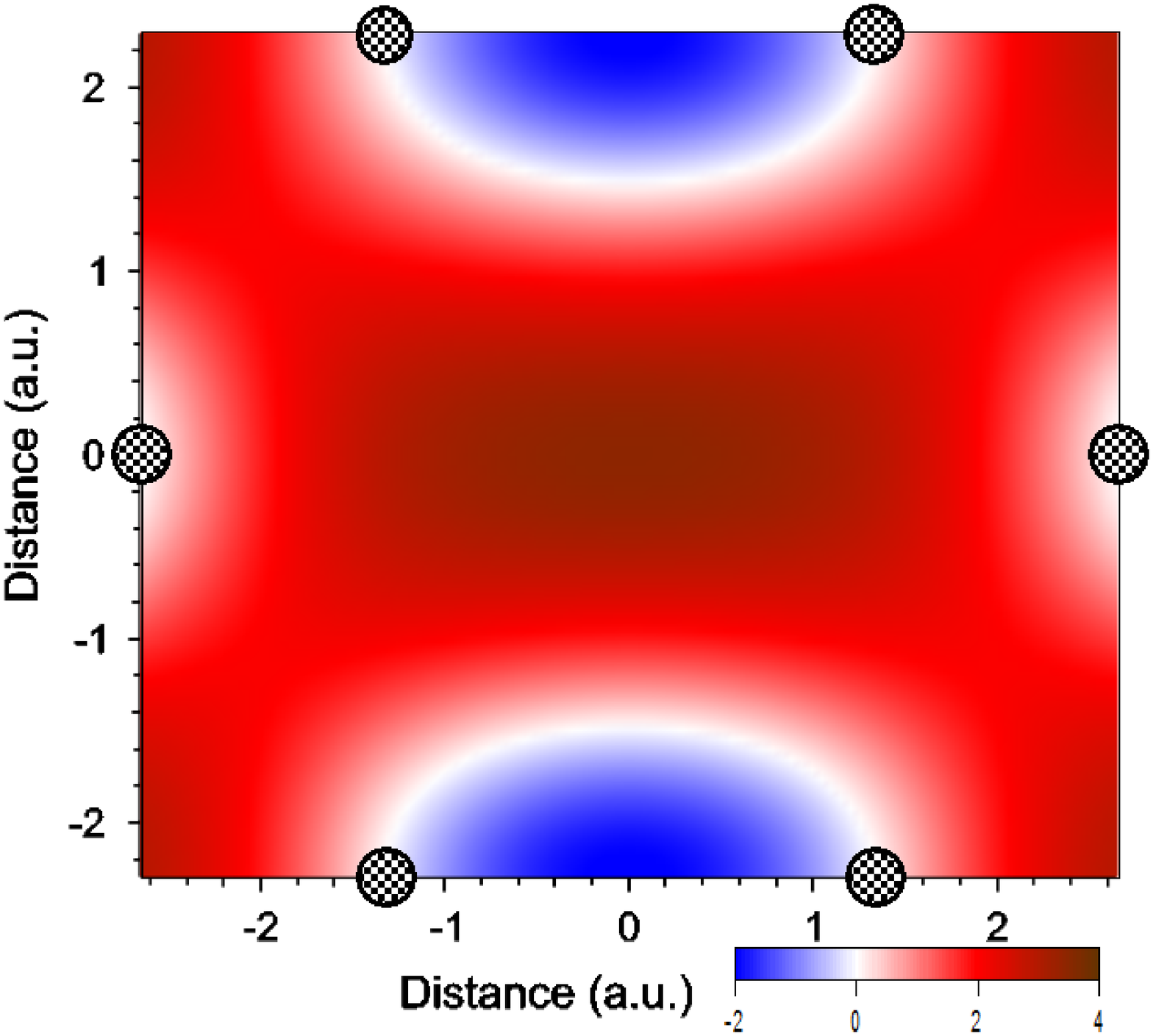}
\end{minipage}
\caption{\label{fig:6k}  (color online)  Real (left) and imaginary (right) parts of the wave function (in arbitrary
units) at the $z = 1$ a.u. plane for the sixth lowest band at $K$ point. }
\end{figure}

\begin{figure}[h]
%\centering
\hskip -.5cm
  \begin{minipage}[b]{0.25\textwidth}
  \includegraphics[width= 1\textwidth]{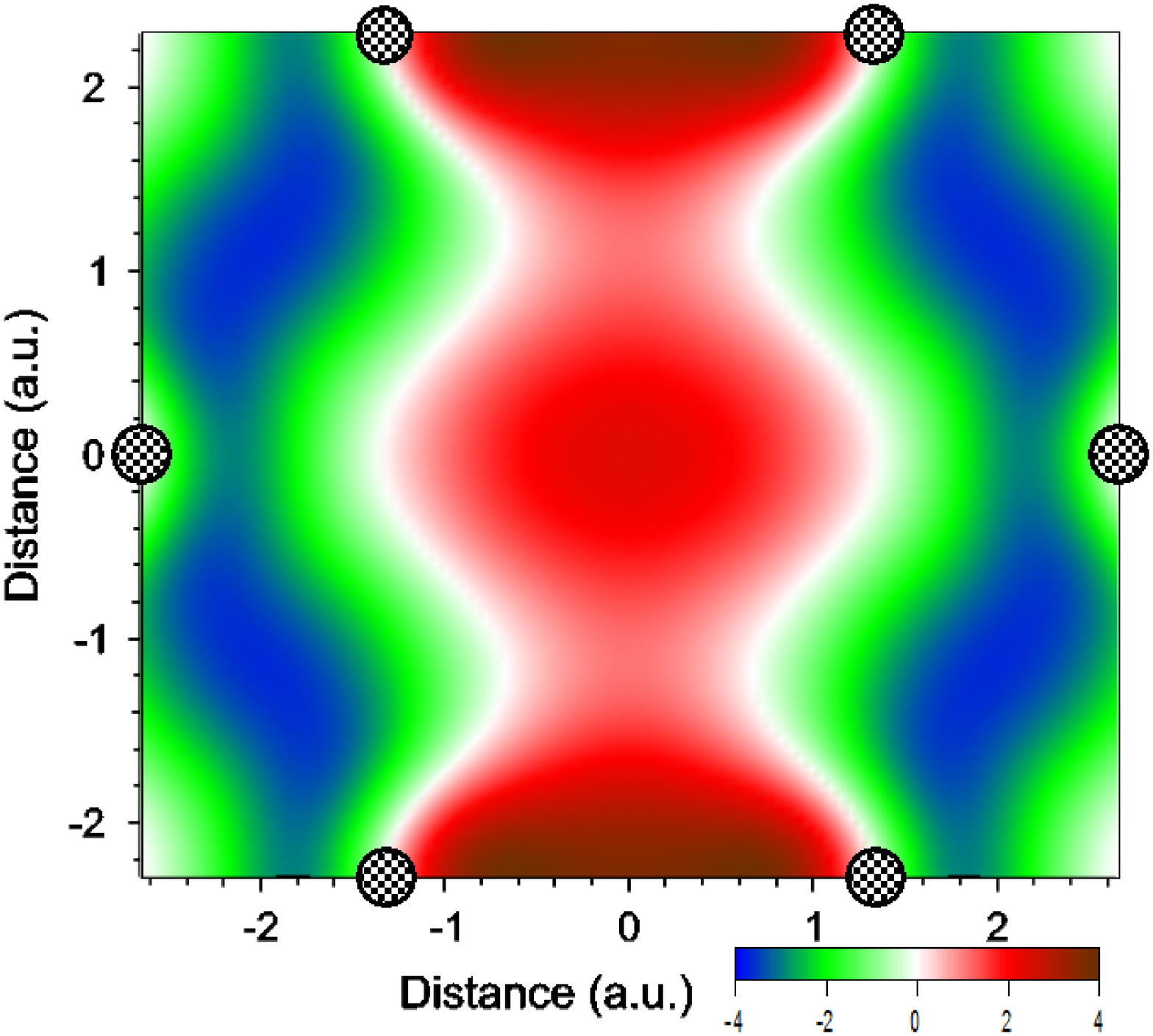}
\end{minipage}
  \hfill
  \begin{minipage}[b]{0.25\textwidth}
\includegraphics[width= 1\textwidth]{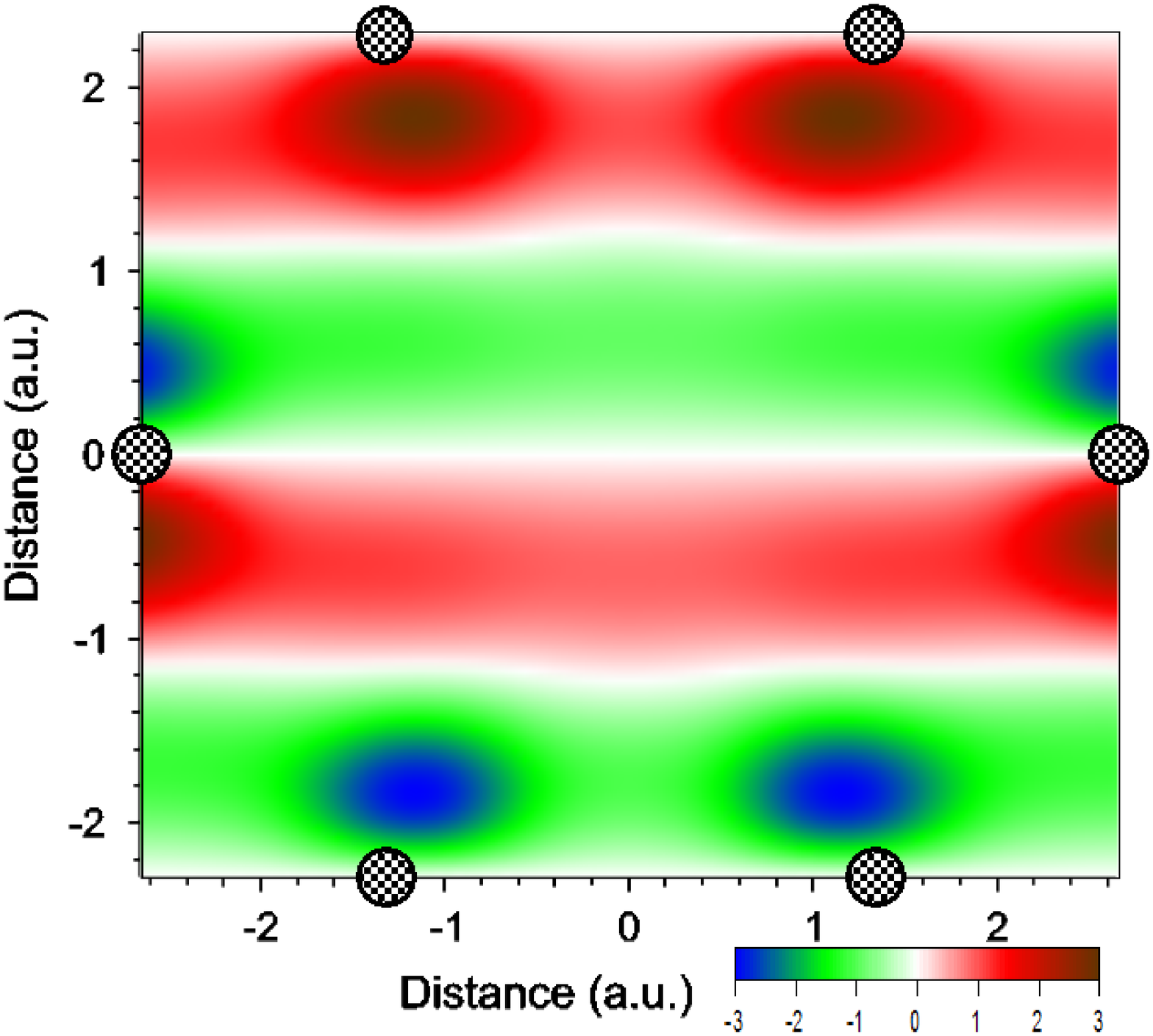}
\end{minipage}
\caption{\label{fig:9k}  (color online)  Real (left) and imaginary (right) parts of the wave function (in arbitrary
units) at the $z = 0$ plane for the ninth lowest band at $K$ point. }
\end{figure}

On Figs. \ref{fig:3k} -\ref{fig:9k} we present the results
of the calculations of the wave functions  of the non degenerate bands at the point $K$. The wave functions of the $\sigma$ bands are plotted at the plane $z=0$. For the  $\pi$ band,  the wave function is identically equal to zero at the $z=0$ plane, so we plotted the wave function at the plane $z=1$ a.u.
All the Figures are consistent with the assignments made in Fig. \ref{fig:bands}.

%\begin{figure}[h]
%\centering
%\hskip -.5cm
%  \begin{minipage}[b]{0.25\textwidth}
%  \includegraphics[width= 1\textwidth]{K90_point_n=1_Re_Z=0.eps}
%\end{minipage}
%  \hfill
%  \begin{minipage}[b]{0.25\textwidth}
%\includegraphics[width= 1\textwidth]{K90_point_n=1_Im_Z=0.eps}
%\end{minipage}
%\caption{\label{fig:1}  (color online)  Real (left) and imaginary (right) parts of the wave function (in arbitrary
%units) in the $z = 0$ plane for the lowest band at $K$ point. }
%\end{figure}
%\begin{figure}[h]
%\centering
%\hskip -.5cm
%  \begin{minipage}[b]{0.25\textwidth}
%  \includegraphics[width= 1\textwidth]{K90_point_n=4_Re_Z=1au.eps}
%\end{minipage}
%  \hfill
%  \begin{minipage}[b]{0.25\textwidth}
%\includegraphics[width= 1\textwidth]{K90_point_n=4_Im_Z=1au.eps}
%\end{minipage}
%\caption{\label{fig:1}  (color online)  Real (left) and imaginary (right) parts of the wave function (in arbitrary
%units) at the $z = 1$ a.u. plane for the forth lowest band at $K$ point. }
%\end{figure}
%\begin{figure}[h]
%\centering
%\hskip -.5cm
%  \begin{minipage}[b]{0.25\textwidth}
%  \includegraphics[width= 1\textwidth]{K90_point_n=10_Re_Z=1au.eps}
%\end{minipage}
%  \hfill
%  \begin{minipage}[b]{0.25\textwidth}
%\includegraphics[width= 1\textwidth]{K90_point_n=10_Im_Z=1au.eps}
%\end{minipage}
%\caption{\label{fig:1}  (color online)  Real (left) and imaginary (right) parts of the wave function (in arbitrary
%units) at the $z = 1$ a.u. plane for the tenth lowest band at $K$ point. }
%\end{figure}

In our previous publication \cite{kogan1} we presented simple TBM interpretation of the degeneracy of the $\pi$ bands at the point $K$ ($E''$ representation). The general TBM  Hamiltonian for the $\pi$ bands is
\begin{eqnarray}
\label{hamz}
H =-
\left(\begin{array}{cc}
E_p+\sum_{\bf a} t'({\bf a})e^{i{\bf k\cdot a}} & \sum_{\bf a}t({\bf a}+{\bf \delta})e^{i{\bf k\cdot}({\bf a}+{\bf \delta})}\\
\sum_{\bf a}t^*({\bf a}+{\bf \delta})e^{-i{\bf k\cdot}({\bf a}+{\bf \delta})} & E_p+ \sum_{\bf a}t'({\bf a})e^{i{\bf k\cdot a}} \end{array}\right),\nonumber\\
\end{eqnarray}
where $E_p$ is the energy of an isolated $|p_z>$ orbital, ${\bf a}$ is an arbitrary  lattice vector.
The structure of graphene
can be seen as a triangular lattice with a basis of two
atoms per unit cell, displaced from each other by any one (fixed) vector $\delta$ connecting two sites of different sub-lattices.
From Ref. \cite{neto} we get that
${\bf a}$  is a linear combination of
${\bf a}_1=\frac{a}{2}\left(3,\sqrt{3}\right)$, ${\bf a}_2=\frac{a}{2}\left(3,-\sqrt{3}\right)$ and ${\bf \delta}$ can be chosen as
${\bf \delta}=-a\left(1,0\right)$. Also
$K=\frac{2\pi}{3a}\left(1,\frac{1}{\sqrt{3}}\right)$.

Consider, for example, three terms in   the sum $\sum_{\bf a}t({\bf a}+{\bf \delta})e^{i{\bf k\cdot}({\bf a}+{\bf \delta})}$  corresponding to hopping between the nearest neighbours. The hopping integral $t({\bf a}+{\bf \delta})$ is the same for all three terms;
the multiplier $e^{i{\bf k\cdot}({\bf a}+{\bf \delta})}$  takes  three values: $1,e^{2\pi i/3},e^{-2\pi i/3}$, $\epsilon =e^{2\pi i/3}$. Hence, due to the identity
$1+e^{2\pi i/3}+e^{-2\pi i/3}=0$, the non-diagonal terms in the Hamiltonian (\ref{hamz}) disappear;
hence degeneracy of the bands.

Similar interpretation can be supplied for the degeneracy of the $\sigma$ bands at the K point.
The reasoning from the previous paragraph can be repeated verbatim for the $E'$ representation realized by
$\psi^A_{s}$ and $\psi^B_{s}$ functions.
The TBM analysis of the   representations realized by
$\psi^{A,B}_{p_{x,y},}$ functions is a bit more complicated.

Let us start defining a convenient for our purpose basis for the $E$ representation of the group $C_{3v}$ realized by $|p_{x,y}>$ orbitals:
\begin{eqnarray}
|p_L>=|p_x>+\epsilon \hat{T}|p_x>+\epsilon^2 \hat{T}^2|p_x>\nonumber\\
|p_R>=|p_x>+\epsilon^2 \hat{T}|p_x>+\epsilon \hat{T}^2|p_x>,
\end{eqnarray}
where $\hat{T}$ is an operator of rotation by the angle $2\pi/3$.
If we graphically present $|p_x>$ orbital by a line in the $x$-direction, then the basis vectors can be presented as two stars.

\setlength{\unitlength}{.4cm}
\begin{picture}(6,6)
\thicklines

\put(2, 3){ \line(-1,-2){1}}
\put(2, 3){ \line(-1,2){1}}
\put(2, 3){ \line(1,0){2.3}}
\put(2,4){$\epsilon$}
\put(3.5,3.2){$1$}
\put(2,1.5){$\epsilon^2$}

\put(8, 3){ \line(-1,-2){1}}
\put(8, 3){ \line(-1,2){1}}
\put(8, 3){ \line(1,0){2.3}}
\put(9.5,3.2){$1$}
\put(8,4){$\epsilon^2$}
\put(8,1.5){$\epsilon$}

\end{picture}

\noindent
It is obvious  that
rotation of each basis vector by $2\pi/3$ just multiplies it by $\epsilon$ ($\epsilon^2$).

Returning to our original representation we chose the basis vectors as
\begin{eqnarray}
\begin{array}{l}
\psi_1\equiv \psi^A_{L}=\sum_{{\bf R}_A} e^{i{\bf K\cdot R}_A}|p_L>_{{\bf R}_A}\\
\psi_2\equiv \psi^A_{R}=\sum_{{\bf R}_A} e^{i{\bf k\cdot R}_A}|p_R>_{{\bf R}_A}\\
\psi_3\equiv \psi^B_{L}=\sum_{{\bf R}_B} e^{i{\bf k\cdot R}_B}|p_R>_{{\bf R}_B}\\
\psi_4\equiv \psi^B_{R}=\sum_{{\bf R}_B} e^{i{\bf k\cdot R}_B}|p_R>_{{\bf R}_B},
\end{array}
\end{eqnarray}
where $|\dots>_{\bf R}$ means an orbital centered at the site ${\bf R}$.
The Hamiltonian matrix in such basis has the form
\begin{eqnarray}
\label{ham}
H=\left(\begin{array}{cccc} 0 & \dots & 0 & \dots \\ \dots  &  0  & \dots & 0 \\ 0 & \dots & 0 & \dots \\
\dots & 0 & \dots & 0 \\ \end{array}\right),
\end{eqnarray}
where we have shifted energy to $H_{11}+E_p$ ($E_p$ is the energy of an isolated $|p>$ orbital).
The elements indicated by three dots are of no interest to us. The origin of zeros on the diagonal of the  matrix is obvious.
To understand presence of non-diagonal zeros (in a chosen basis) consider, for example, the three terms in   the $H_{13}$ matrix element  corresponding to hopping between the nearest neighbours. They can be graphically presented as

\setlength{\unitlength}{.3cm}
\begin{picture}(12,16)
\thicklines

\put(13, 8){ \line(-1,-2){1}}
\put(13, 8){ \line(-1,2){1}}
\put(13, 8){ \line(1,0){2.3}}
\put(14.5,8.2){$1$}
\put(13,9){$\epsilon$}
\put(13,6.5){$\epsilon^2$}

\put(18, 8){ \line(-1,-2){1}}
\put(18, 8){ \line(-1,2){1}}
\put(18, 8){ \line(1,0){2.3}}
\put(19.5,8.2){$1$}
\put(18,9){$\epsilon$}
\put(18,6.5){$\epsilon^2$}

\put(10.5, 13){ \line(-1,-2){1}}
\put(10.5, 13){ \line(-1,2){1}}
\put(10.5, 13){ \line(1,0){2.3}}
\put(12,13.2){$\epsilon$}
\put(10.5,14){$\epsilon^2$}
\put(10.5,11.5){$1$}

\put(10.5, 3){ \line(-1,-2){1}}
\put(10.5, 3){ \line(-1,2){1}}
\put(10.5, 3){ \line(1,0){2.3}}
\put(12,3.2){$\epsilon^2$}
\put(10.5,4){$1$}
\put(10.5,1.5){$\epsilon$}

\end{picture}

\noindent
Again, the identity $1+e^{2\pi i/3}+e^{-2\pi i/3}=0$ leads to the disappearance of the matrix element. The structure of the matrix (\ref{ham}) shows that its eigenvalues can be grouped into pairs with the the same modulus and the opposite signs. On the other hand, this structure shows that the product of all eigenvalues is equal to zero. Hence degeneracy of the bands at the energy $0$ (that is at the energy $H_{11}+E_p$).

\subsection{$M$ - point}

Finally  consider the point $M$. The functions $\psi_{\bf M}^j$ realize $A_g+B_{3u}$ representation of the group $D_{2h}$. The orbitals $|s>$, $|p_z>$, $|p_x>$ and $|p_y>$ realize $A_g$, $B_{1u}$, $B_{3u}$ and $B_{2u}$ representations respectively.
Thus the identity
\begin{eqnarray}
\label{mmm}
(A_g+B_{3u})\times B_{1u}=B_{1u}+B_{2g}
\end{eqnarray}
shows the symmetry of the  two $\pi$ bands at the point $M$; the identity
\begin{eqnarray}
\label{mm4}
(A_g+B_{3u})\times B_{2u}=B_{2u}+B_{1g}
\end{eqnarray}
shows the symmetry of the  two $\sigma$ bands constructed from $|p_y>$ orbitals, and the identities
\begin{eqnarray}
\label{mm2}
\begin{array}{l}
(A_g+B_{3u})\times B_{3u}=B_{3u}+A_{g}\\
(A_g+B_{3u})\times A_g=A_g+B_{3u}
\end{array}
\end{eqnarray}
show that there are two $A_g$ and two $B_{3u}$ bands  constructed from $|p_x>$ and $|s>$ orbitals.

From analyzing the TBM \cite{neto}  we come to the conclusion   that the band $B_{1u}$
lies above the band $B_{2g}$. From Eq. (\ref{mm2})   we come to the conclusion   that, if
  overlapping between $|p_x>$ orbitals is stronger than  that between $|s>$ orbitals, the lowest band at the point $M$ realizes  $B_{3u}$ reprezentation.

\begin{figure}[h]
%\centering
\hskip -.5cm
  \begin{minipage}[b]{0.25\textwidth}
  \includegraphics[width= 1\textwidth]{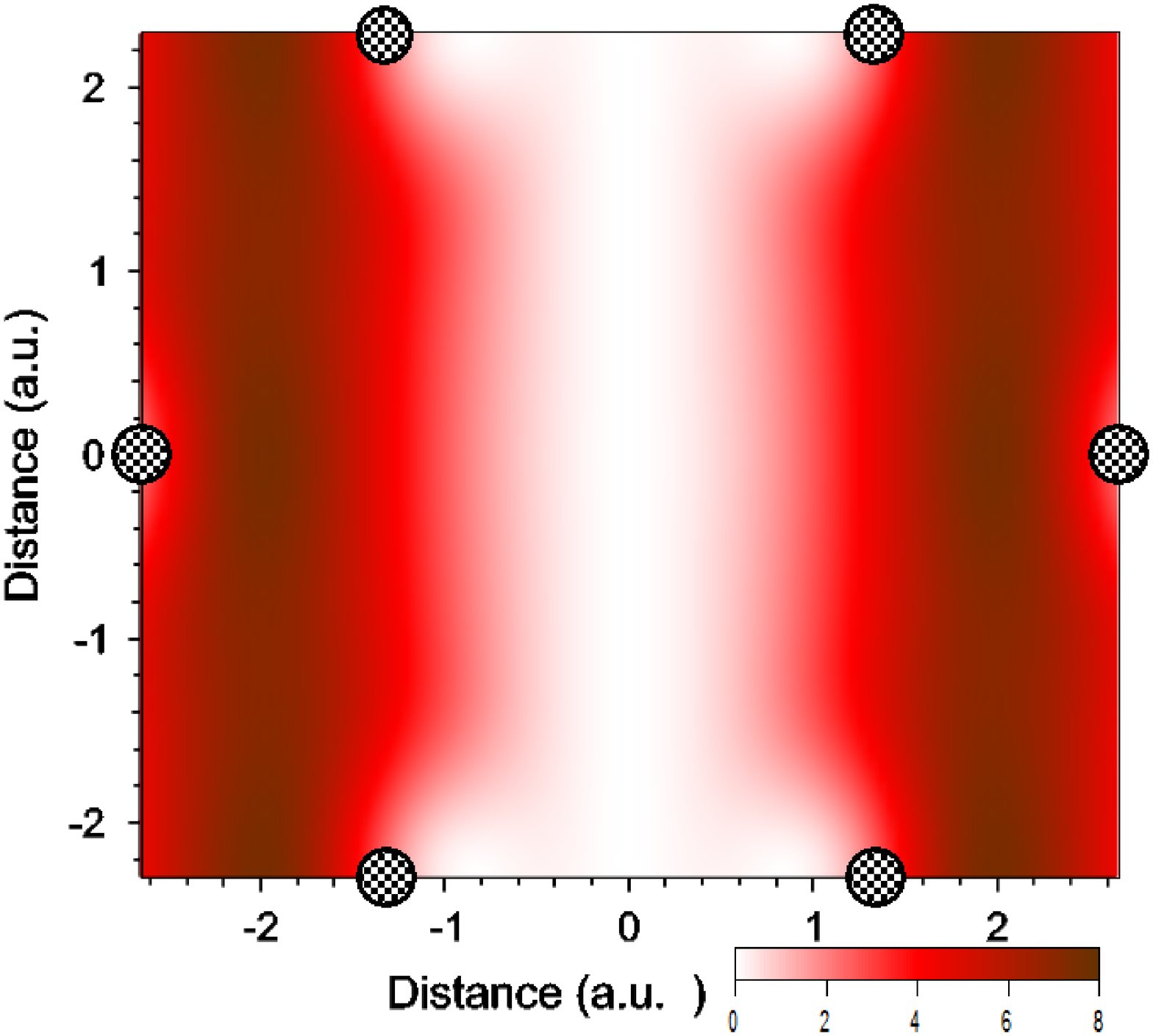}
\end{minipage}
  \hfill
  \begin{minipage}[b]{0.25\textwidth}
\includegraphics[width= 1\textwidth]{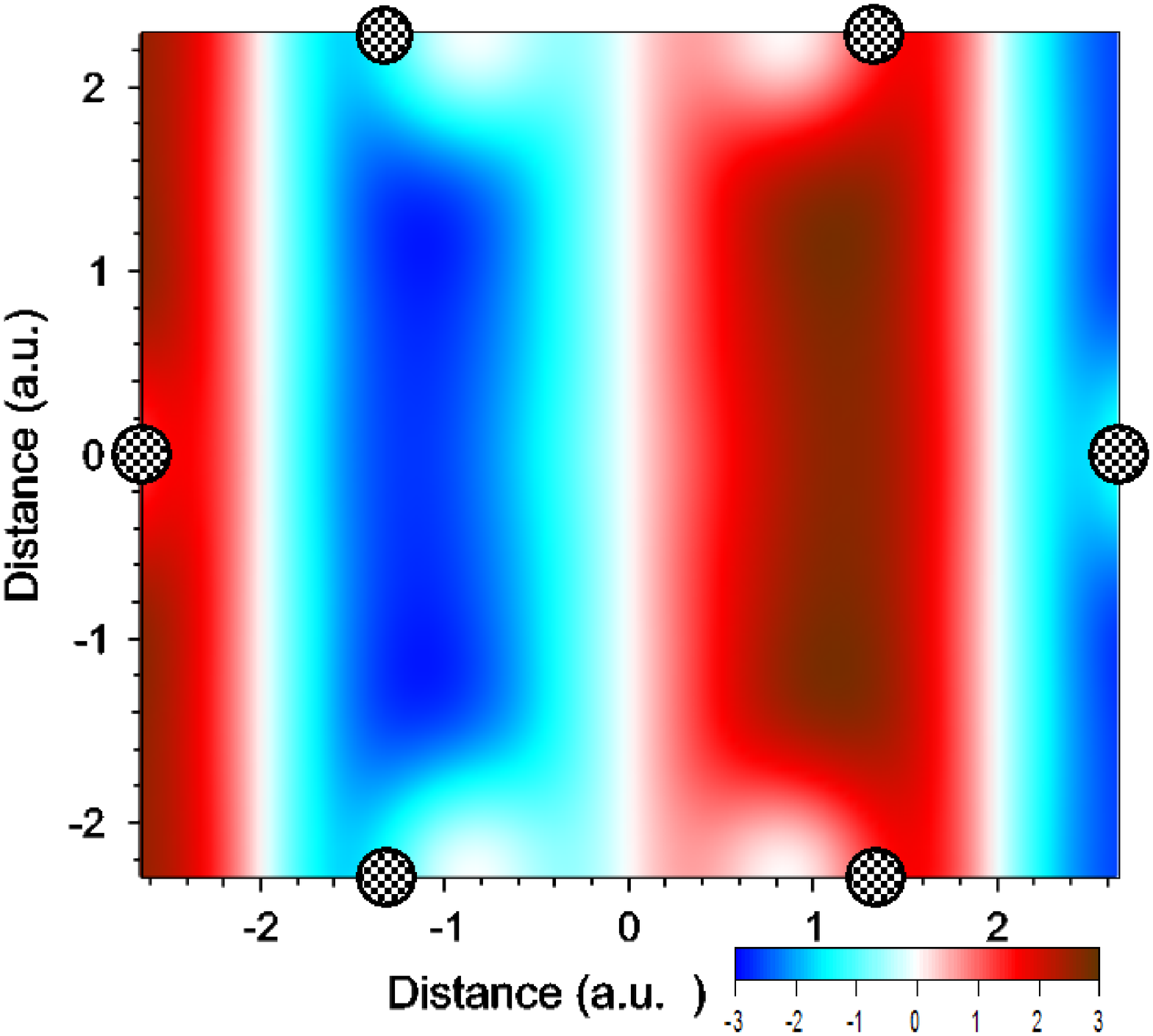}
\end{minipage}
\caption{\label{fig:1}  (color online)  Real (left) and imaginary (right) parts of the wave function (in arbitrary
units) at the $z = 0$ plane for the lowest band at $M$ point. }
\end{figure}

\begin{figure}[h]
%\centering
\hskip -.5cm
  \begin{minipage}[b]{0.25\textwidth}
  \includegraphics[width= \textwidth]{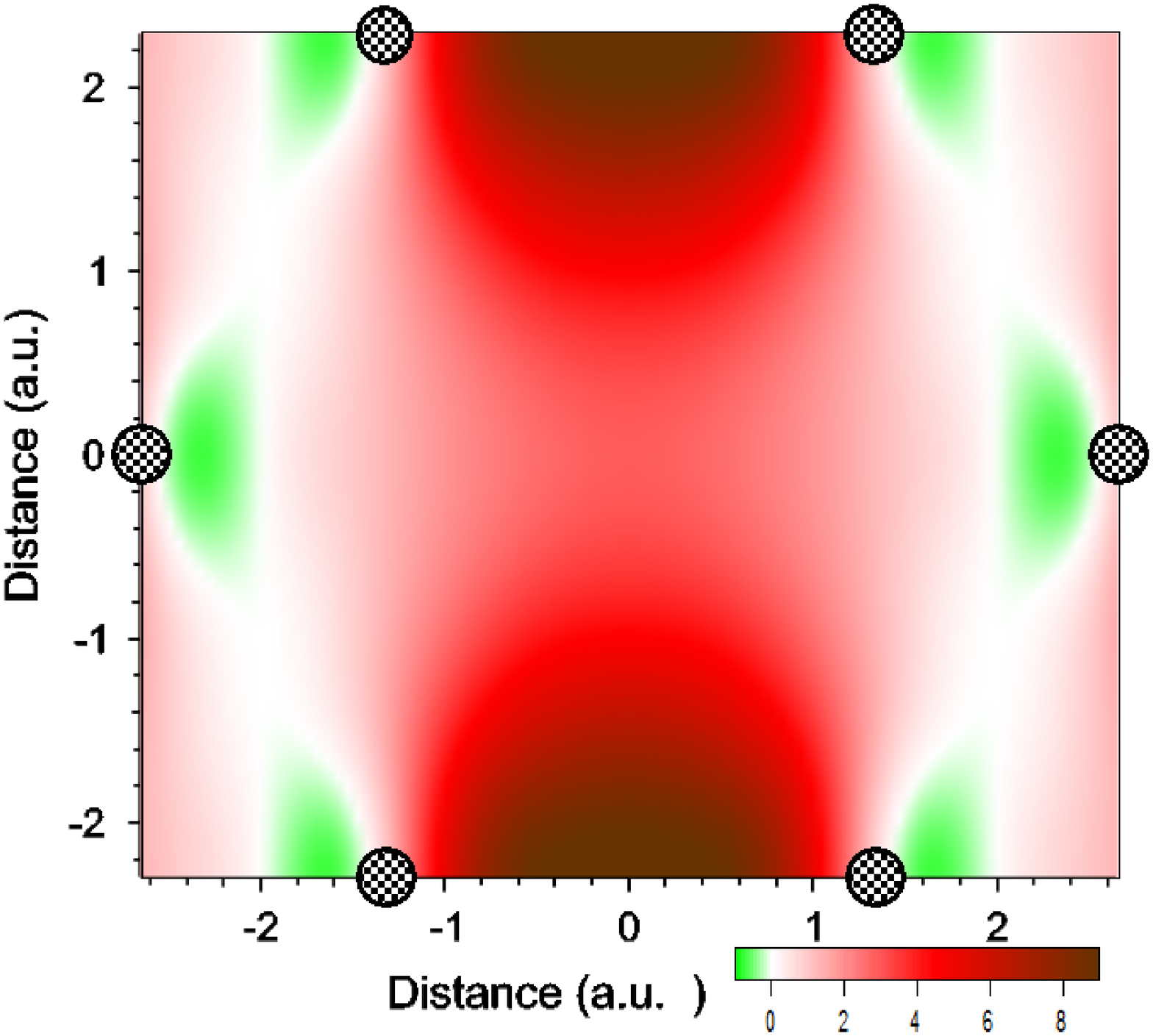}
\end{minipage}
  \hfill
  \begin{minipage}[b]{0.25\textwidth}
\includegraphics[width= \textwidth]{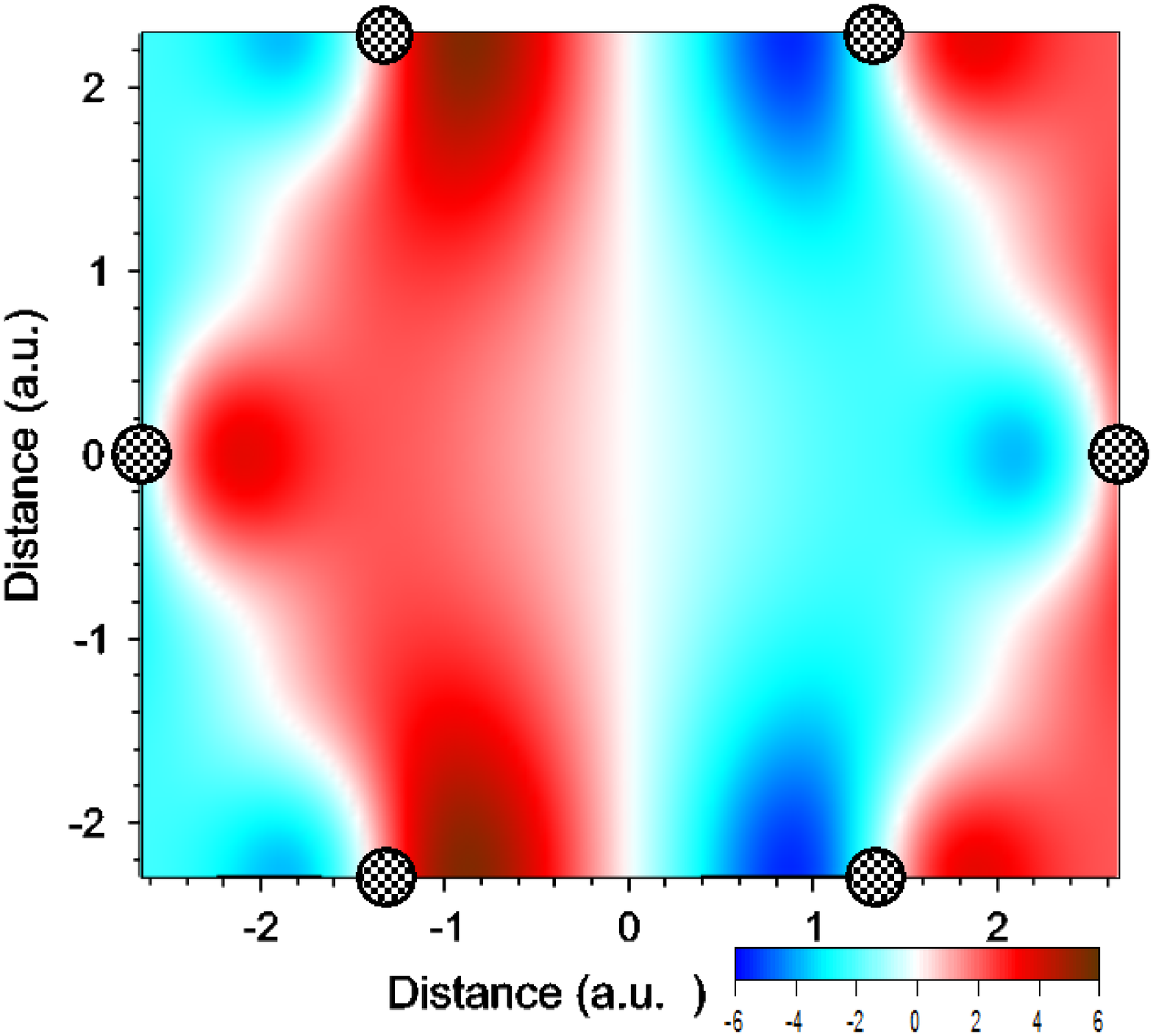}
\end{minipage}
\caption{\label{fig:2}  (color online)  Real (left) and imaginary (right) parts of the wave function (in arbitrary
units) at the $z = 0$ plane for the second lowest band at $M$ point. }
\end{figure}

\begin{figure}[h]
%\centering
\hskip -.5cm
  \begin{minipage}[b]{0.25\textwidth}
  \includegraphics[width= \textwidth]{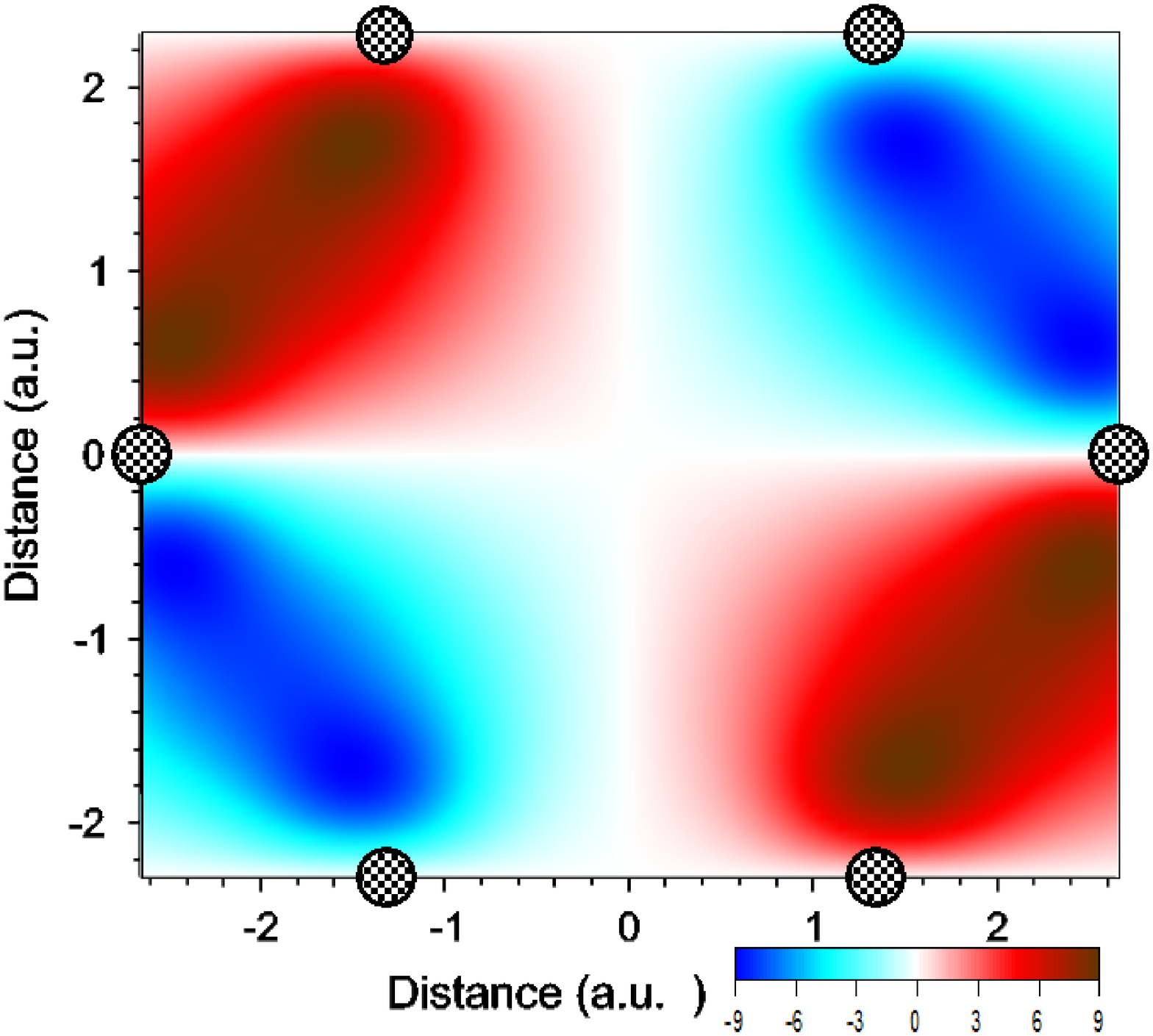}
\end{minipage}
  \hfill
  \begin{minipage}[b]{0.25\textwidth}
\includegraphics[width= \textwidth]{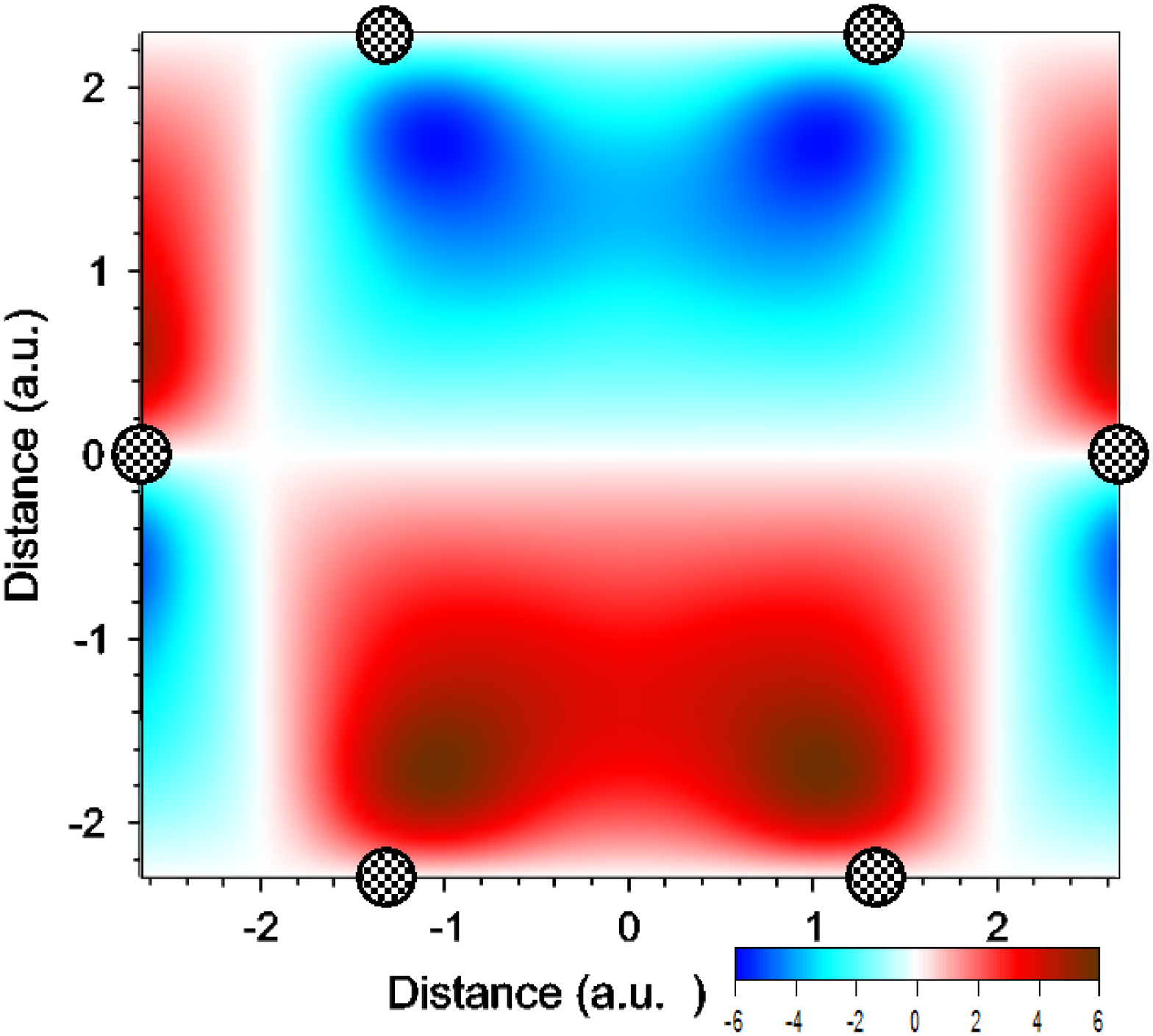}
\end{minipage}
\caption{\label{fig:3}  (color online)  Real (left) and imaginary (right) parts of the wave function (in arbitrary
units) at the $z = 0$ plane for the third lowest band at $M$ point. }
\end{figure}

\begin{figure}[h]
%\centering
\hskip -.5cm
  \begin{minipage}[b]{0.25\textwidth}
  \includegraphics[width= \textwidth]{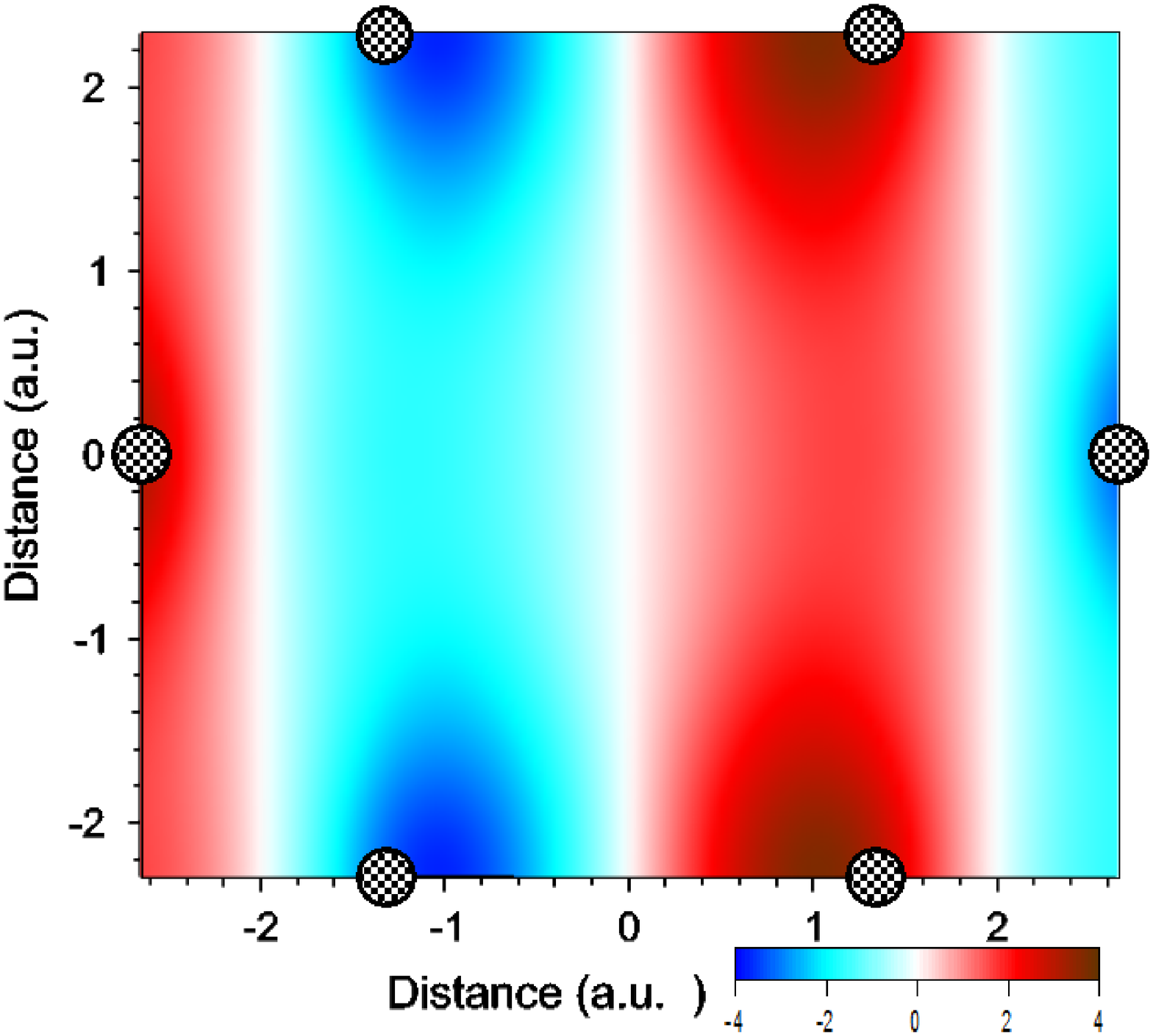}
\end{minipage}
  \hfill
  \begin{minipage}[b]{0.25\textwidth}
\includegraphics[width= \textwidth]{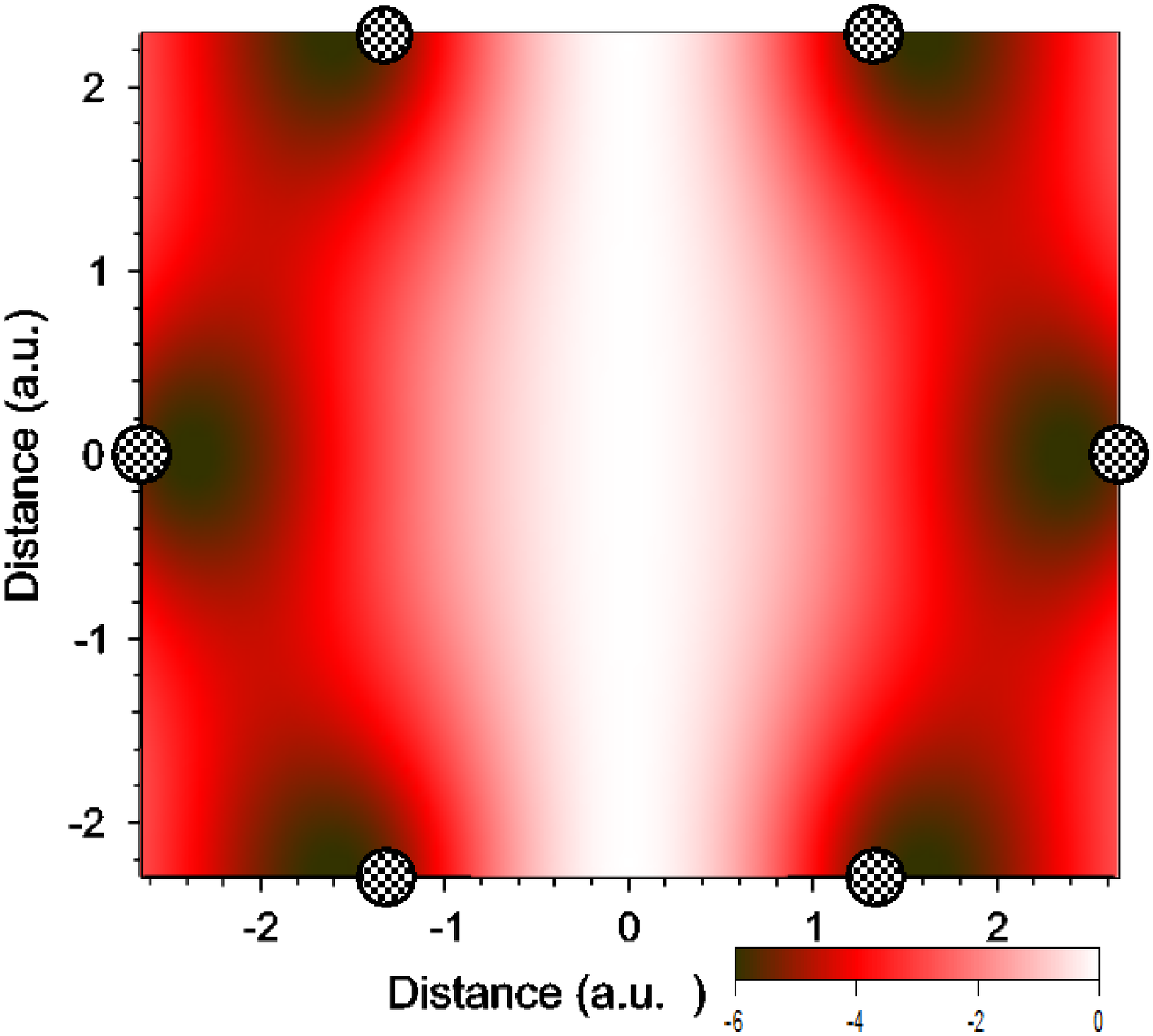}
\end{minipage}
\caption{\label{fig:4}  (color online)  Real (left) and imaginary (right) parts of the wave function (in arbitrary
units) at the $z = 1$ a.u. plane for the forth lowest band at $M$ point. }
\end{figure}

\begin{figure}[h]
%\centering
\hskip -.5cm
  \begin{minipage}[b]{0.25\textwidth}
  \includegraphics[width= \textwidth]{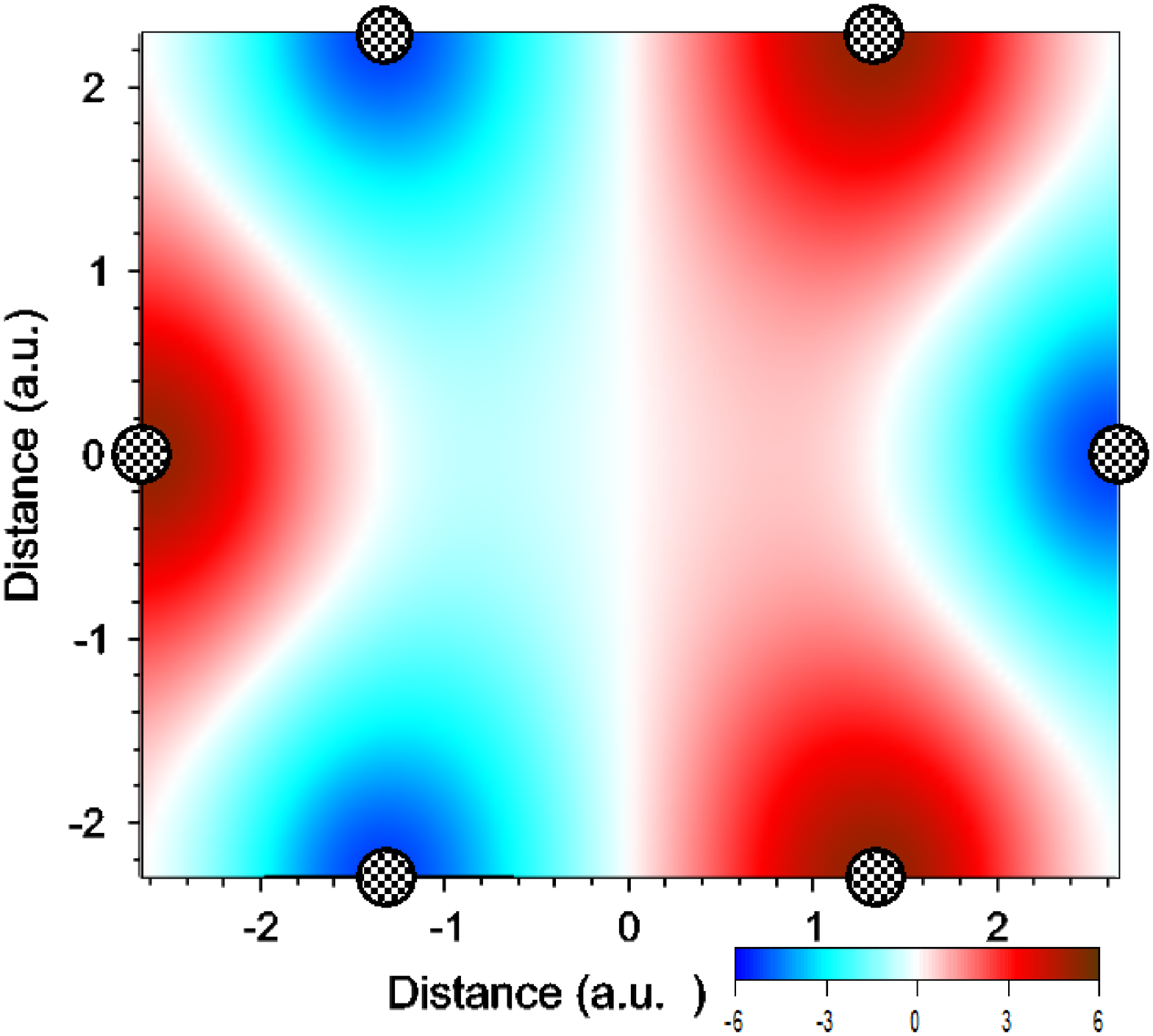}
\end{minipage}
  \hfill
  \begin{minipage}[b]{0.25\textwidth}
\includegraphics[width= \textwidth]{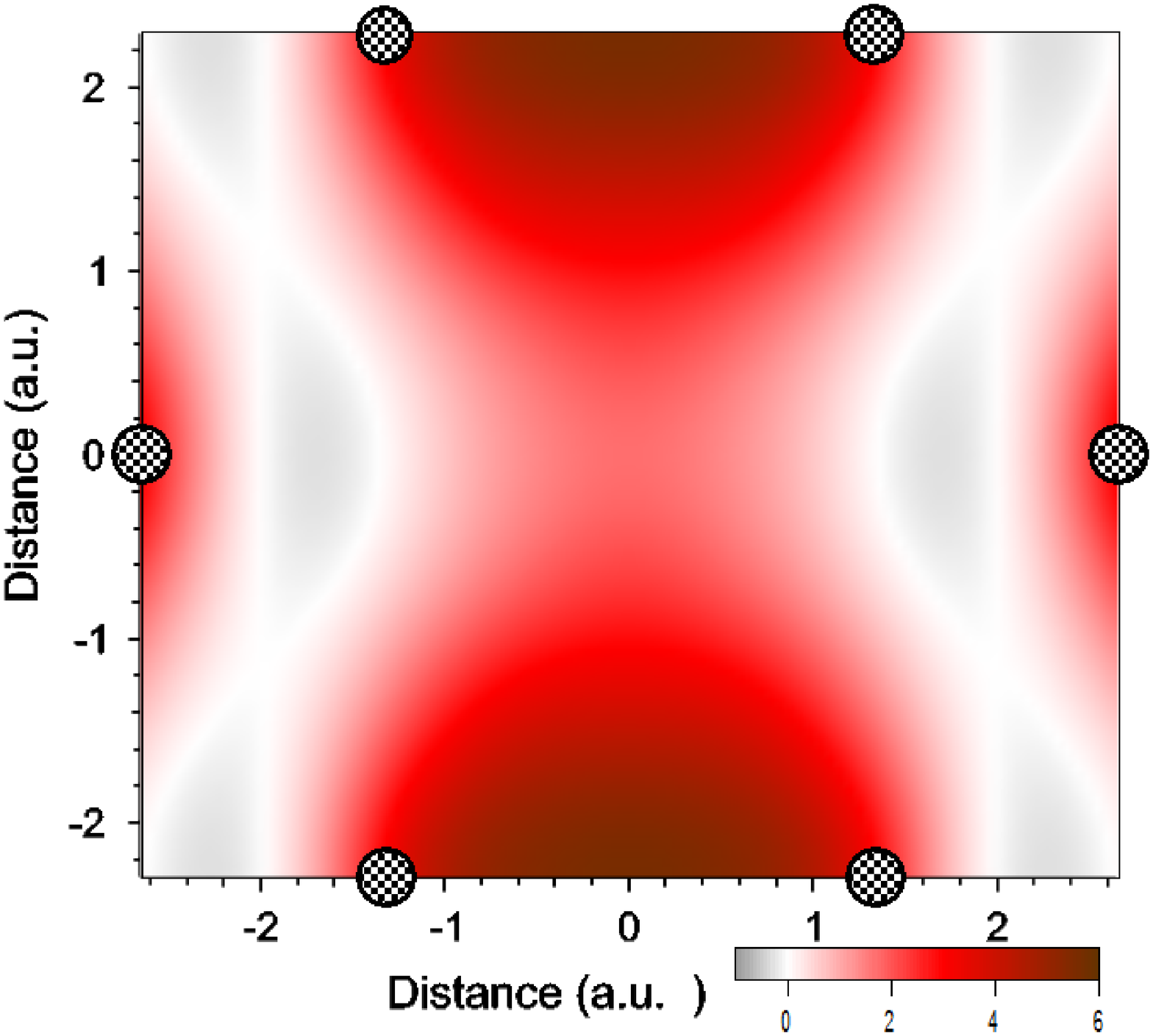}
\end{minipage}
\caption{\label{fig:5}  (color online)  Real (left) and imaginary (right) parts of the wave function (in arbitrary
units) at the $z = 1$ a.u. plane for the fifth lowest band at $M$ point. }
\end{figure}

On Figs. \ref{fig:1} -\ref{fig:5} we present the results
of the calculations of the wave functions  of the occupied  and the lowest unoccupied bands at the point $M$. The wave functions of the $\sigma$ bands are plotted at the plane $z=0$. For the  $\pi$ bands,  the wave function is identically equal to zero at the $z=0$ plane, so we plotted the wave function at the plane $z=1$ a.u.

For the lowest (at the point $M$) band the wave function
 is equal to zero along the $y$ - axis,
which corresponds to the representation $B_{3u}$. (Because the wave function is antisymmetric with respect to reflection, it should be
equal to zero at the axis of reflection.) The wave function of the next band is different from zero everywhere at the plane,
which is consistent with the representation $A_g$. The wave function of the third band is equal to zero at the $x$ -axis,
which corresponds to the representation $B_{2u}$.

For the forth band  the wave function  is equal to zero along the $y$ - axis,
which corresponds to the representation $B_{2g}$. The wave function of the first unoccupied  band is different from zero everywhere at the plane $z=1$ a.u.,
which is consistent with the representation $B_{1u}$.


\subsection{$\Gamma-K-M$ - lines}

The symmetry of the band(s) at the symmetry point determines the
symmetry of the band(s) at the symmetry lines containing this point
\cite{landau,thomsen}. At the line $\Gamma-K-M$ the functions
$\phi_{\bf k}^{A,B}$ realize  $A_1+B_1$ representation of the group
$C_{2v}$. The orbital $|p_z>$ realizes $B_2$ representation. From
the identities $B_2\times A_1=B_2$, $B_2\times B_1=A_2$  we come to
the conclusion that the lower band should realize $B_2$
representation. At the line $\Gamma-M$ the functions $\phi_{\bf
k}^{A,B}$ realize twice $A_1$ representation of the group $C_{2v}$.
The orbital $|p_z>$ generalizes $B_2$ representation. From the
identity $B_2\times A_1=B_2$  we obtain that both $\pi$ bands
realize $B_2$ representation.

\begin{acknowledgments}

The authors are grateful   for the useful discussions to P. D. Esquinazi, M. I. Katsnelson,  A. I. Lichtenstein, M. Saito, V. U. Nazarov, N. S. Pavlov, O. Rader, L. M. Sandratskii,  A. Varykhalov, and S. Yunoki.
They are also grateful to P. D. Esquinazi for  bringing to their attention Ref. \cite{hund}.

\end{acknowledgments}

\section{Appendix}

For convenience of the reader in  we reproduce character tables for the groups used in the paper \cite{landau}
(Tables \ref{table:d85} and \ref{table:d2}).

\begin{table}
%\caption{Character table}
\begin{tabular}{|l|l|rrrr|}
\hline
\hline
 $C_{2v}$ &  & $E$ & $C_2$ & $\sigma_v$ &  $\sigma_v'$ \\
& $D_2$ & $E$ & $C_2^z$ & $C_2^y$ &  $C_2^x$ \\
\hline
$A_1;z$ & $A$ & $1$ & 1 & 1 & 1 \\
$B_2;y$ & $B_3;x$ & $1$ & $-1$  & $-1$ & 1    \\
$A_2$ & $B_1;z$ & $1$ & 1 & $-1$ & $-1$ \\
$B_1;x$ & $B_2;y$ & $1$ & $-1$ & 1 & $-1$     \\
\hline
\hline
\end{tabular}
\caption{Character table for irreducible representations of $C_{2v}$ and  $D_2$ point groups}
\label{table:d85}
\end{table}

\begin{table}
%\caption{Character table}
\begin{tabular}{|l|l|rrrrrr|}
\hline
\hline
 $D_6$ &  & $E$ & $C_2$ & $2C_3$ & $2C_6$ & $3U_2$ & $3U_2'$ \\
 & $D_{3h}$   & $E$ & $\sigma$ & $2C_3$ & $2S_3$ & $3U_2$ & $3\sigma_v$ \\\hline
 $A_{1}$ & $A_1'$ & 1 & 1 & 1 & 1 & 1 & 1 \\
 $A_{2}$ & $A_2'$ & 1 & 1 & 1 & 1 & $-1$ & $-1$ \\
 $B_{1}$ & $A_1''$ & 1 & $-1$ & 1 & $-1$ & 1 & $-1$ \\
 $B_{2}$  &  $A_2''$ &1 & $-1$ & 1 & $-1$ & $-1$ & 1 \\
 $E_{2}$  & $E'$ & 2 & 2 & $-1$ & $-1$ & 0 & 0 \\
 $E_{1}$ & $E''$  & 2 & $-2$  & $-1$ & 1 & 0 & 0 \\
\hline
\hline
\end{tabular}
\caption{Character table for irreducible representations of  $D_6$ and $D_{3h}$ point groups}
\label{table:d2}
\end{table}
The characters of the groups $D_{2h}$ and $D_{6h}$ are obtained using the relations $D_{2h}=D_2\times C_i$ and $D_{6h}=D_6\times C_i$,
where $C_i$ is the inversion group.

\eject

\end{document}